# Features of Explainability: How users understand counterfactual and causal explanations for categorical and continuous features in XAI

## GRETA WARREN[*]

Insight SFI Centre for Data Analytics, School of Computer Science, University College Dublin, Dublin, Ireland, greta.warren@ucdconnect.ie

## MARK T. KEANE

Insight SFI Centre for Data Analytics, VistaMilk SFI Research Centre, School of Computer Science, University College Dublin, Dublin, Ireland, mark.keane@ucd.ie

## RUTH M.J. BYRNE

School of Psychology and Institute of Neuroscience, Trinity College Dublin, University of Dublin, Dublin, Ireland, rmbyrne@tcd.ie

Counterfactual explanations are increasingly used to address interpretability, recourse, and bias in AI decisions. However, we do not know how well counterfactual explanations help users to understand a system's decisions, since no large-scale user studies have compared their efficacy to other sorts of explanations such as causal explanations (which have a longer track record of use in rule-based and decision-tree models). It is also unknown whether counterfactual explanations are equally effective for categorical as for continuous features, although current methods assume they do. Hence, in a controlled user study with 127 volunteer participants, we tested the effects of counterfactual and causal explanations on the objective accuracy of users' predictions of the decisions made by a simple AI system, and participants' subjective judgments of satisfaction and trust in the explanations. We discovered a dissociation between objective and subjective measures: counterfactual explanations elicit higher accuracy of predictions than no-explanation control descriptions but no higher accuracy than causal explanations, yet counterfactual explanations elicit greater satisfaction and trust than causal explanations. We also found that users understand explanations referring to categorical features more readily than those referring to continuous features. We discuss the implications of these findings for current and future counterfactual methods in XAI.

## 1 INTRODUCTION

The use of automated decision making in computer programs that impact people's everyday lives has led to rising concerns about the fairness, transparency, and trustworthiness of Artificial Intelligence (AI; see e.g., [54,56]). These concerns have created renewed interest in, and an urgency about, tackling the problem of eXplainable AI (XAI). Recently, counterfactual explanations have been advanced as a promising solution to the XAI problem because of their compliance with data protection regulations, such as, the EU's General Data Protection Regulation (GDPR; see [64]), their potential to support algorithmic recourse [30], and their psychological importance in explanation [9,51]. The prototypical XAI scenario for counterfactuals involves the explanation of an automated decision when a banking app refuses a customer's loan application; on querying the decision, the customer is told "if you had asked for a lower loan of $10,000, your application would have been approved". On the face of it, people appear to readily understand these counterfactual explanations and they also offer people possible recourse to change the decision's outcome (e.g., by lowering their loan request). Although there is now a substantial XAI literature on counterfactuals, because of a lack of user studies we know very little about *how* people understand these counterfactual explanations of AI decisions, and *which* aspects of counterfactual methods are critical to their use in XAI. In this paper, we address this gap in the literature with a statistically well-powered and psychologically well-controlled user study (N=127) examining how different explanations impact people's understanding of automated decisions. We test explanations of automated decisions about blood alcohol content and legal limits for driving using counterfactual explanations (e.g., 'if John had drunk 3 units instead of 5 units, he would have been under the limit'), compared to causal explanations (e.g., 'John was over the limit because he drank 5 units), and descriptions ('John was over the limit'). The study examines not only the effects of explanations but also the effects of different types of features – categorical features (gender, stomach-fullness) and continuous features (units, duration of drinking, body weight). It includes objective measures of the accuracy of participants' understanding of the automated decision, and subjective measures of their satisfaction and trust in the system and its decisions. In the remainder of this introduction, we introduce the relevant related work in this area on counterfactual explanations (see 1.1), causal explanations (see 1.2), and how feature-types have been handled in XAI systems (see 1.3), before outlining the current experiment (see 1.4).

---

[*] Corresponding author

**1.1 Counterfactual Explanations**

In recent years, XAI research on the use of counterfactuals has exploded, with over 100 distinct computational methods proposed in the literature (for reviews see [29,31,62]). These various techniques argue for different approaches to counterfactual generation; some advance optimisation techniques [52,64], others emphasise the use of causal models [28,29], distributional analyses [12,25,35] or the importance of instances [20,32,47]. These alternative proposals are typically motivated by claims that the method in question generates "good" counterfactuals for end-users; for instance, that the counterfactuals are psychologically "good" because they are proximal [64], plausible [28], actionable [45], sparse [32] or diverse [52]. However, most of these claims are based on intuition rather than on empirical evidence. A recent review found that just 21% of 117 papers on counterfactual explanation included any user-testing, and fewer (only ~7%) tested specific properties of the method proposed [31]. This state of affairs raises the possibility that many of these techniques contain functions with little or no psychological validity, that may have no practical benefit to people in real-life applications [1,37].

Consider what we *have* learned from the few user studies on counterfactuals in XAI. Most user studies test whether counterfactual explanations impact people's responses relative to no-explanation controls or some other explanation strategy (e.g., use of example-based explanations or rule-based explanations [63]). These studies assess explanation quality using "objective" measures (e.g., user predictive accuracy) and/or "subjective" measures (e.g., user judgments of trust, satisfaction, preference). In philosophical and psychological research, explanations are understood to be designed to change people's understanding of the world, events or phenomena [33,39,43]. In XAI, this definition has been conceptualised to mean that explanation should improve people's understanding of the AI system, the domain involved in the task and/or their performance on the target task (e.g., [23,34]). An explanation is effective, therefore, if people *objectively* perform better on a task involving the AI system by, for example, being faster, more accurate, or by being able to predict what the system might do next [20,36,40,44,63]. Concretely, if a person with diabetes is using an app to estimate their blood sugar levels for insulin treatments, ideally the system's predictions would help them better understand their condition in the future; for example, their predictions of their own blood sugar levels should improve, when the app's help is not available. However, XAI research has also focused on whether explanations work *subjectively*; that is, whether the explanation improves people's trust or satisfaction in the AI system, or whether the explanation makes people "feel better" about their interaction with the system in the task. Hence, many user studies ask people to rate their trust and satisfaction with the system [23], the fairness of its decisions [3,13], their preference for one explanation over another [17,59], or to judge their own understanding of the system [40,44]. These subjective measures shed a different, but important, light on the efficacy of explanations. Indeed, an explanation strategy that elicits positive objective *and* subjective outcomes may well be the holy grail of XAI. However, if there is a disassociation between the objective and subjective impacts of an explanation strategy then there may be cause for concern. Notably, if an explanation strategy has no objective impact on understanding but is subjectively preferred by users, then concerns about its ethical use could arise. Emerging evidence of such disassociations in some task contexts where counterfactual explanations have been used raises the worrying prospect that users may feel more confident about an app's advice without learning anything about the domain or how the app works, thus developing an *inappropriate trust* in it.

With respect to *objective measures*, there are a handful of studies showing mixed support for the use of counterfactual explanations in XAI. *What-if, why-not, how-to* and *why* explanations have been found to improve performance in prediction and diagnosis tasks relative to no-explanation controls, however counterfactual (*what-if*) explanations did not improve performance appreciably more than other explanation options [40]. Visual counterfactual explanations were shown to increase classification accuracy relative to no-explanation controls in a small sample of users [20]. However, neither contrastive rule-based nor example-based explanations for a simulated blood-sugar prediction app performed better than controls in people's ability to predict correct insulin dosages [63]. Interestingly, one study reported that counterfactual tasks, in which users were asked if a system's recommendation would change given a perturbation of some input feature, elicited longer response times and greater judgments of difficulty, and lower accuracy than simulation tasks in which users were asked to predict the recommendation based on the input features [36]. Similarly, people have been found to be less accurate when asked to produce a counterfactual change for an instance than when predicting an outcome from the features [44]. These findings are consistent with the proposal that counterfactuals require people to consider multiple possibilities, to compare reality to the suggested alternative and to infer causal relations, as is often reported in the psychological literature [8,19,57]; and that they aid users in reasoning about past decisions, but require cognitive effort and resources [2,7,46]. However, some caution should be exercised in generalising from the small collection of XAI user studies on counterfactual explanations, given the diversity of tasks, domains and experimental designs; some do not involve controls, and many others use too few test items or very small numbers of participants to be confident about the findings reported.

With respect to *subjective measures*, there are a number of studies showing that people have positive assessments of counterfactual explanations. Sensitivity-based explanations, describing how input to a system would have to change to alter the outcome, are judged as more appropriate and fair than example-based explanations [3]. Further supportive evidence suggests that these sorts of explanations increase people's fairness judgments of the system and improve their ability to spot case-specific fairness issues in comparison to other explanation strategies (such as case-based, demographic-based and influence-based ones [13]). Providing contrastive explanations in a sales-forecasting domain has been found to increase self-reported understanding of the system's decisions [44]. When asked to judge counterfactual explanations with varying degrees of sparsity along dimensions of concreteness, relevance and coherence, participants appeared to prefer explanations with 3-4 feature differences [18]; also, model-generated counterfactual instances were judged to be more realistic and typical than instances generated using values drawn at random from the training data, akin to judgements of real-world instances along these dimensions [16].

However, two studies have shown dissociations between objective and subjective measures. Users shown contrastive rule-based explanations self-reported better understanding of the system's decision than no-explanation controls, however neither of these groups, nor users shown contrastive example-based explanations, showed any improvement in accuracy for predicting what the system might do, and tended to follow the system's advice, even when incorrect [63]. A similar disconnect between objective and subjective evaluation metrics was found for systematically increasing the complexity of a system's causal rules; although users' response times and judgments of difficulty also increased, little effect of complexity was observed on task accuracy [36]. Clearly, there are many issues to be resolved about the relationships and dependencies between objective measures and subjective judgments [6,48]. In XAI, studies asking users how well they understand a system's decisions or how satisfying they find an explanation, may not accurately reflect the true explanatory power of different sorts of explanations, particularly given people's propensity to overestimate their understanding of complex causal mechanisms [58]. In the current study, we assess the extent to which counterfactual and causal explanations increase people's understanding using measures that are objective (accuracy in predicting what the system will do) and subjective (i.e., judgments of trust and satisfaction).

### 1.2 Causal Explanations

The present work compares counterfactual explanations to those using causal rules, as the latter are a long-standing explanation strategy in AI. In philosophical and psychological research, there is consensus that everyday explanations often invoke some notion of cause and effect [33], framed as exchanges of causal information about an event [41][1]. In AI, such causal explanations are often cast as IF-THEN rules (e.g., in expert systems such as MYCIN [6, 4] or decision trees [24,36]). In XAI, it is commonly claimed that such rule-based explanations are inherently interpretable, although some have pointed out that this claim may not be the case [14,42]. One study reports that when users are given causal decision sets for a system, they achieve high accuracy in a prediction task, along with low response times and low subjective difficulty judgments; however in a counterfactual task using the same decision sets, performance was poorer on all three of these measures [36]. Another study found that contrastive rule-based explanations were effective in helping users identify the decisive feature in a system's decision and increased people's sense of understanding of the system; indeed contrastive rule based explanations were more effective than contrastive example-based explanations [63]. These findings suggest that causal rules may be as good as, if not sometimes better than, counterfactual explanations in some task contexts. In the psychological literature, it has been found that when people are asked to reflect on an imagined negative event, they spontaneously generate twice as many causal explanations as counterfactual thoughts about it [50], consistent with the proposal that causal explanations do not require people to think about multiple possibilities in the way that counterfactual explanations do. Indeed, eye-tracking data shows that people look at only a single image on-screen when they hear a causal assertion, whereas they look at multiple images when they hear a counterfactual [55]. Since it has also been shown that people make difficult inferences from counterfactuals more readily than they do from their factual counterparts [10], it is plausible that counterfactuals' evocation of multiple possibilities may help users consider an AI system's decision more deeply. Given the clear importance of these two explanation options – causal and counterfactual strategies – both are compared to one another in the present study. Considering that their appeal to the purportedly contrastive nature of explanation [51] is one of the main arguments for their use in XAI, and given the psychological evidence that counterfactuals are understood by thinking about more possibilities than causal explanations, we predict that counterfactual explanations will aid users in understanding the system's decisions in comparison to causal explanations, while both will outperform mere descriptions of an outcome.

---

[1] Indeed, causal explanations and counterfactuals have long been viewed as being intertwined in complex ways [23, 40, 66], though psychologically they clearly differ in significant ways. For instance, when creating causal explanations the adduced causes may be necessary or merely sufficient for an outcome to occur, whereas when people construct counterfactuals they tend to undo causes that are necessary for the outcome to occur [47].

### 1.3 Feature-Types in Explanations

The present study also examines the role of different feature-types in explanations, which have been emphasised in various counterfactual techniques; although the types of features examined here have been largely overlooked in the XAI literature. Many counterfactual methods distinguish between the types of features to be used in explanations; it makes sense to use features that are *mutable* rather than *immutable* [26] (e.g., being told to "reduce your age to get a loan" is not very useful). Furthermore, proponents argue that the features used in the counterfactual should be *causally important* [21,29,47] and/or *actionable* [11,28,61]; a counterfactual explanation proposing to reduce the size of the requested loan is more actionable and therefore better than one telling the customer to modify a long-standing, bad credit-rating. However, psychologically, perhaps there are more fundamental feature properties to be considered, such as whether people can understand continuous (such as income or credit-score) or categorical (such as race or gender) features equally well. Psychological studies have long shown that people do not spontaneously make changes to continuous variables such as time or speed, e.g., when they imagine how an accident could have been avoided [27]. This distinction between continuous and categorical features is important, because if people are less likely to manipulate continuous features or have difficulties understanding counterfactuals about them, then the potential causal importance or actionability of such features is moot. For example, in algorithmic recourse, people may better understand counterfactual advice that says, "you need to change your credit-score from bad to good" rather than saying "you need to increase your credit score from 3.4 to 4.6". To date, the differential impacts of categorical and continuous feature-types has not been considered in counterfactual methods for XAI. Most counterfactual generation methods treat categorical feature values as being undifferentiated from continuous ones (e.g., DiCE [52] applies one-hot encoding to categorical feature values). In the present study, we examine people's understanding of counterfactual explanations for different feature-types (continuous versus categorical), predicting that explanations focusing on categorical features will be more readily understood, leading to greater predictive accuracy based on these features.

### 1.4 Outline of Current Study

The study tests the impact of counterfactual versus causal explanations, and continuous versus categorical features, on users' accuracy of understanding and subjective evaluation of a simulated AI system designed to predict blood alcohol content and legal limits. Participants are shown predictions by the system for different instances, with explanations (e.g. "If Mary had weighed 80kg instead of 75kg, she would have been under the limit."). The study consists of two phases (i) a *training phase* in which participants are asked to predict the system's decision (i.e. an individual being over or under the legal blood alcohol content threshold to drive a car), and are provided with feedback on the system's predictions and with explanations for each decision, and (ii) a *testing phase,* in which they are asked to predict outcomes for a different set of test instances, this time with no feedback nor any explanations. In the training phase, participants consider the system's predictions and learn about the blood alcohol content domain with the help of the explanations, to determine whether this experience *objectively* improves their understanding of the domain. The testing phase objectively measures their developed understanding of the system using the accuracy of their predictions. Users' subjective evaluations are also recorded using judgments of satisfaction and trust.

## 2 THE TASK: PREDICTING LEGAL LIMITS FOR DRIVING

Participants were presented with the output of a simulated AI system presented as an app, designed to predict whether someone is over the legal blood alcohol content limit to drive. The system relies on a commonly-used approximate method, the Widmark equation [65,67], that uses five key features for blood alcohol content with the limit threshold being set at 0.08% alcohol per 100ml of blood. This formula was used to generate a dataset of instances for normally-distributed values of the feature-set, from which the study's materials were drawn (N=2000).

In the experimental task, participants were instructed that they would be testing a new app, *SafeLimit*, designed to inform people whether or not they are over the legal limit to drive, from five features: *units* of alcohol consumed by the person, *weight* (in kg), *duration* of drinking period (in minutes), *gender* (male/female) and *stomach-fullness* (full/empty). The experiment consisted of two phases. In the training phase, participants were shown examples of tabular data for different individuals, and asked to make a judgment about whether each individual was under or over the limit on each screen. Participants selected one of three options: 'Over the limit', 'Under the limit', or 'Don't know' by clicking the corresponding on-screen button. The order of these options was randomised, to ensure that participants did not merely click on the same button-order each time. After giving their response, feedback was given on the next page, with the correct answer highlighted using a green tick-mark, and the incorrect answer (if selected) highlighted using a red X-mark (see Figures 1 and 2). Above the answer options, participants were also shown an explanation, corresponding to the experimental condition being tested. Figures 1 and 2 show sample materials used in the counterfactual and causal conditions, respectively. Note, that in both conditions the explanations draw attention to a key feature (e.g., the units drunk) as being critical to the prediction made. In all the study's conditions, a balanced set of instances were used, with eight items for each of the five features presented. Upon completing the training phase, participants began the testing phase (see Figure 3). Again, participants were shown instances referring to individuals (different to those in the training phase), and asked to judge if the individual was over or under the legal limit to drive. After submitting their response, no feedback or explanation was given, and participants moved on to the next trial. For each instance, participants were asked to consider a specific feature in making their prediction; for instance, "Given this person's WEIGHT, please make a judgment about their blood alcohol level." Again, in this phase, a balanced set of instances were used, with eight items for each of the five features presented.

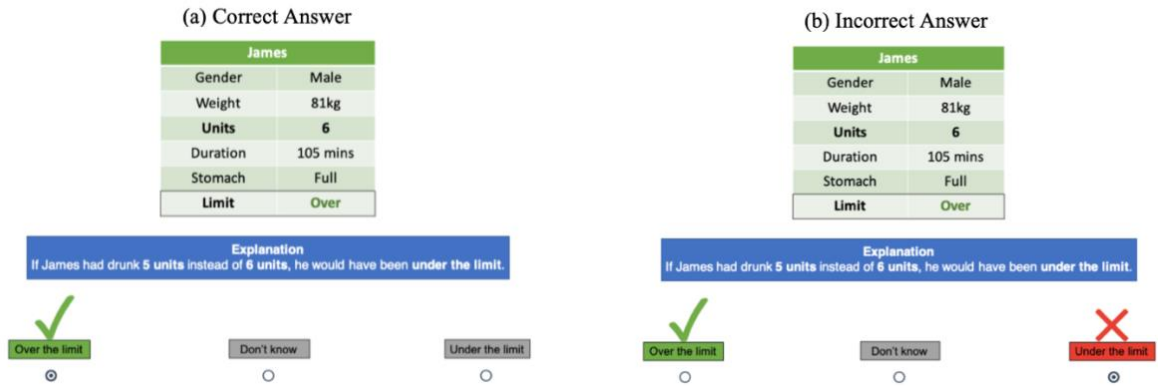

Figure 1: Feedback for (a) Correct Answer and (b) Incorrect Answer in the Counterfactual condition of the study

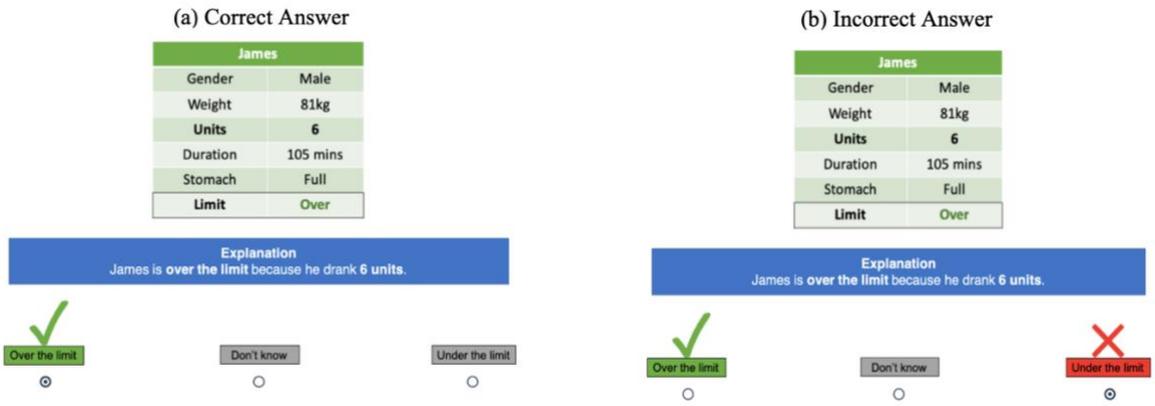

Figure 2: Feedback for (a) Correct Answer and (b) Incorrect Answer in the Causal condition of the study

The objective measure of performance in both phases of the study was accuracy (i.e., correct predictions made by participants compared to those of the system). The subjective measures were explanation satisfaction and trust in the system, assessed using the DARPA project's Explanation Satisfaction and Trust scales [23] respectively (see Appendix A.1). To assess engagement with the task, participants completed four attention checks at random intervals throughout the experiment, and were asked to recall the 5 features used by the app by selecting them from a list of 10 options at the end of the session.

Figure 3: Example of prediction task in testing phase

## 3 METHOD

We compare the impact of counterfactual and causal explanations, to descriptions of the system's decisions as a control condition, on the predictions people make about the *SafeLimit* app's decisions. Participants were assigned in fixed order to one of three groups (counterfactual, causal, control) and completed the experiment, consisting of (i) a *training phase* in which they made predictions and were given feedback with explanations or descriptions and (ii) a *testing phase* where they made predictions with no feedback and no explanations (for all groups). Hence, any observed differences in accuracy in the testing phase should reflect people's understanding of the AI system based on their experiences in the training phase, which differed only in the nature of the explanation (or control description) provided. People were presented with 40 items in each phase, which were systematically varied in terms of the five features used with balanced occurrence (i.e., 8 instances for each feature). Explanation satisfaction and trust in the system were measured following the training and testing phases. Our primary predictions are: (i) explanations should improve accuracy, that is, performance in the training phase should be more accurate than performance in the testing phase, (ii) counterfactual explanations should improve accuracy more than causal explanations, as they are potentially more informative, (iii) predictions about categorical features should be more accurate than predictions about continuous features, if people find the former less complex than the latter, and (iv) counterfactual explanations will be judged as more satisfying and trustworthy than causal explanations, given previous studies showing that they are often subjectively preferred over other explanations.

### 3.1 Participants & Design

The participants, crowdsourced using the Prolific platform (N=127), were randomly assigned to the three between-participant conditions: counterfactual explanation (n=41), causal explanation (n=43) and control (n=43). These groups consisted of 80 women, 46 men, and one non-binary person aged 18-74 years ($M$=33.54, $SD$=13.15); and were pre-screened to select native English speakers from Ireland, the United Kingdom, the United States, Australia, Canada and New Zealand, who had not participated in previous related studies. The experimental design was a 3 (Explanation: counterfactual, causal, control) x 2 (Task: training vs testing phase) x 5 (Feature: units, duration, gender, weight, stomach-fullness), with repeated measures on the latter two variables. A further 11 participants were excluded from the data analysis, one for giving identical responses for each trial, and 10 who failed more than one attention check or memory check. Before testing, the power analysis with G*Power [15] indicated that 126 participants were required to achieve 90% power for a medium effect size with alpha <.05 for two-tailed tests. Ethics approval for the study was granted by the responsible institution with the reference code LS-E-20-11-Warren-Keane.

## 3.2 Materials

Eighty instances were randomly selected, based on key filters from the 2000-item dataset generated for the blood alcohol content domain (based on stepped increments of a feature's normally-distributed values with realistic upper/lower limits). Specifically, the procedure randomly selected an instance (query case) and incrementally increased or decreased one of the five feature's values until its blood alcohol content value crossed the decision boundary to create a counterfactual case. For the categorical features, *gender* and *stomach-fullness*, the inverse value was assigned, while continuous variables were incremented in steps of 15kg for *weight*, 15 minutes for *duration* and 1 *unit* for alcohol. If the query case could not be perturbed to cross the decision boundary, a different case was randomly selected, and the procedure was re-started. If the perturbation was successful, the instance was selected as a material and its counterfactual was used as the basis for the explanation shown to the counterfactual group. For example, if an instance with *units* = 4 crossed the decision boundary when it was reduced by one unit (to be under rather than over the limit) the counterfactual explanation read "If John had drunk 3 units instead of 4 units, he would have been under the limit". The matched *causal explanation* read "John is over the limit because he drank 4 units", with the *control group* given a description of the outcome (e.g., "John is over the limit"). This selection procedure was performed 16 times for each feature, a total of 80 times, with the further constraint that an equal number of instances were found on either side of the decision boundary (i.e., equal numbers under and over the limit). Each instance was then randomly assigned to one of two sets of materials, each comprising 40 items, again ensuring an equal number of instances were classified as under/over the limit. To avoid any material-specific confounds, the materials presented in the training and testing phases were counterbalanced, so that half of participants in each group saw Set A in the training phase, and Set B in the testing phase, while this was reversed for the other half of participants. After data collection, t-tests verified that there was no effect of material-set order.

## 3.3 Procedure

Participants read detailed instructions about the tasks (see Appendix A.2) and completed one practice trial for each phase of the study before commencing. They then progressed through the presented instances, randomly re-ordered for each participant, within the training and testing phases. After completing both phases, they completed the Explanation Satisfaction and Trust scales. Participants were debriefed and paid £2.61 for their time. The experiment took approximately 28 minutes to complete.

## 4 RESULTS & DISCUSSION

Overall, the results show that providing explanations improves the accuracy of people's predictions, and that categorical features lead to higher accuracy levels than continuous features. Participants' accuracy on categorical features was markedly higher in the testing phase relative to the training phase, whereas their accuracy on continuous features remained at essentially the same levels in both phases (an effect that occurred independently of the explanation type). Participants judged counterfactual explanations to be more satisfying and trustworthy than causal explanations, however with regard to participants' accuracy in predicting the AI system's decisions, counterfactual explanations had only a slightly greater impact than causal explanations. The data for this experiment are publicly available at https://osf.io/wqdtn/.

### 4.1 Analysis of the Accuracy Measure

A 3 (Explanation: counterfactual, causal, control) x 2 (Task: training vs testing) x 5 (Feature: units, duration, gender, weight, stomach fullness) mixed ANOVA with repeated measures on the second two factors was conducted on the proportion of correct answers given by each participant (see Figure 4). A Huynh-Feldt correction was applied to the main effect of Feature and its interactions. Significant main effects were found for Explanation, $F(2,124)=5.63$, $p=.005$, $\eta_p^2=.083$, for Task, $F(1,124)=32.349$, $p<.001$, $\eta_p^2=.207$, and for Feature, $F(3.945, 489.156)=47.599$, $p<.001$, $\eta_p^2=.277$. Task interacted with Feature, $F(4, 496)=7.23$, $p<.001$, $\eta_p^2=.055$[2]. These effects were further examined in post hoc analyses.

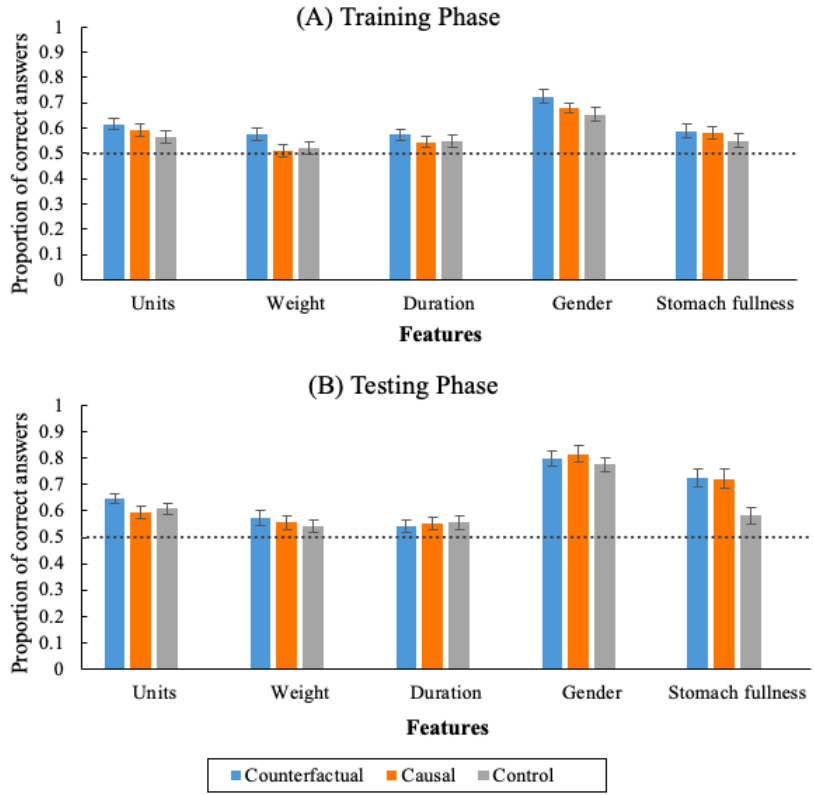

Figure 4: Mean accuracy (proportion of correct answers) across conditions for each feature in the (A) training and (B) testing phases of the study (error bars represent standard error of the mean; dashed line represents chance accuracy).

First, with respect to the main effect of Explanation, post hoc Tukey HSD tests showed that the Counterfactual group ($M=.636$, $SD=.08$) was more accurate than the Control group ($M=.590$, $SD=.08$), $p=.003$, $d=.22$. However, the Causal group ($M=.614$, $SD=.09$) did not differ significantly from the Counterfactual, $p=.245$, or Control groups, $p=.186$. Moreover, there is a reliable trend in increasing accuracy with the following ordering of the groups on their accuracy scores (Page's L(40)=1005.0, $p<.001$): Counterfactual > Causal > Control. These results suggest that providing explanations is better than not providing them, with respect to improving accuracy. They also show, as predicted, that counterfactual explanations have a greater impact than causal explanations relative to controls given no explanations. Note that these effects are seen in both phases of the study (as Explanation does not interact with Task).

---

[2] No other two-way interactions were reliable, neither Explanation with Task, $F(2, 124)=.759$, $p=.47$, nor Explanation with Feature, $F(7.89, 489.156)=1.14$, $p=.335$, nor was the three-way interaction significant, $F(8, 496)=1.215$, $p=.288$.

Second, with respect to the significant Task and Feature variables, and their significant interaction, the decomposition of the interaction revealed that accuracy improves from the training to the testing phase for the categorical features (*gender, stomach-fullness*), but *not* for the continuous features (*units, weight* and *duration*). Post hoc pairwise comparisons with a Bonferroni-corrected alpha of .002 for 25 comparisons showed that participants made more correct responses in the testing phase than the training phase when considering *gender*, $t(126)=5.626$, $p<.001$, $d=.50$, and *stomach-fullness*, $t(126)=4.430$, $p<.001$, $d=.39$, but not *units*, $t(126)=1.350$, $p=.179$, *weight*, $t(126)=-1.209$, $p=.229$, or *duration*, $t(126)=.32$, $p=.75$. The analysis also showed that within each phase of the study, the categorical features produce higher accuracy levels than continuous features, confirming the prediction that people find the former easier to understand than the latter. In the training phase, accuracy for *gender* was significantly higher than accuracy for *units*, $t(126)=4.935$, $p<.001$, $d=.44$, *weight*, $t(126)=6.824$, $p<.001$, $d=.61$, *duration*, $t(126)=6.332$, $p<.001$, $d=.58$, and *stomach-fullness*, $t(126)=5.202$, $p<.001$, $d=.46$, all other features did not differ significantly from each other ($p>.05$ for all comparisons). In the testing phase, similar tests found accuracy to be higher for *gender* than for *units*, $t(126)=8.844$, $p<.001$, $d=.78$, *weight*, $t(126)=10.824$, $p<.001$, $d=.96$, *duration*, $t(126)=10.81$, $p<.001$, $d=.96$ and *stomach-fullness*, $t(126)=4.986$, $p<.001$, $d=.44$. Furthermore, accuracy for *stomach-fullness* was significantly higher than that for *weight*, $t(126)=4.943$, $p<.001$, $d=.44$ and *duration*, $t(126)=4.959$, $p<.001$, $d=.44$, and *units*, $t(126)=2.853$, $p=.005$, although the latter was not significant on the corrected alpha.[3]

Indeed, there is some evidence to suggest that it is the diversity in the range of feature values that may lead to these effects, rather than some abstract ontological status of the feature. If we rank-order each of the features in terms of the number of unique values present in the materials, we find that this rank-ordering predicts the observed trend in accuracy in the testing phase. That is, the rank ordering from highest-to-lowest diversity – *duration* (60 unique values) > *weight* (36 unique values) > *units* (4 unique values) > *stomach-fullness* (2 unique values) = *gender* (2 unique values) – inversely predicts the trend in accuracy: *duration* ($M=.549$) < *weight* ($M=.557$) < *units* ($M=.615$) < *stomach-fullness* ($M=.675$), < *gender* ($M=.796$); Page's L(127)=6256.5, $p<.001$.

**4.2 Analyses of the Subjective Measures: Satisfaction and Trust**

All groups completed the DARPA Explanation Satisfaction and Trust scales after completing the two main phases in the experiment (see Figure 5).

---

[3] Accuracy for *units* was significantly higher than when *weight*, $t(126)=3.152$, $p=.002$, $d=.28$ and *duration*, $t(126)=3.539$, $p=.001$, $d=.31$ were considered. Accuracy for *weight* and *duration* did not differ, $t(126)=.385$, $p=.701$.

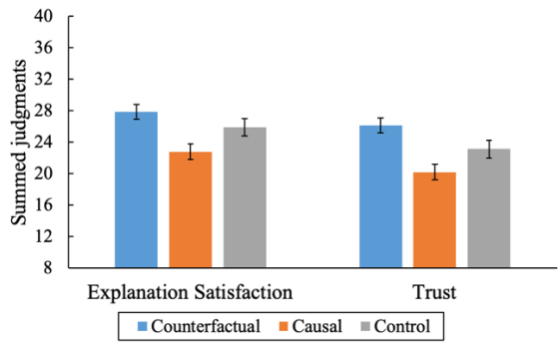

Figure 5: Summed judgments for Explanation Satisfaction and Trust scales (error bars represent standard error of the mean).

*Satisfaction Measure.* A one-way ANOVA was carried out on the summed judgments for the Explanation Satisfaction scale to examine group differences in satisfaction levels for the explanations provided (see Figure 6). Significant differences between the three groups were identified $F(2, 126)=6.104$, $p=.003$, $\eta_p^2=.09$. Post hoc Tukey HSD tests showed that the counterfactual group ($M=27.83$, $SD=6.12$) gave significantly higher satisfaction judgments than the causal group ($M=22.79$, $SD=6.63$), $p=.002$, $d=0.76$. The control group ($M=25.86$, $SD=7.19$) did not differ significantly from either the counterfactual ($p=.369$) or the causal ($p=.087$) groups. A reliable trend was identified when rank-ordering judgments for each item in the order Counterfactual > Control > Causal, Page's L(8)=111.0, $p<.001$, suggesting that counterfactual explanations were marginally more satisfying than descriptions, while the latter were preferred to causal explanations. These results indicate that people are less satisfied with causal explanations, relative to counterfactual explanations or even none at all.

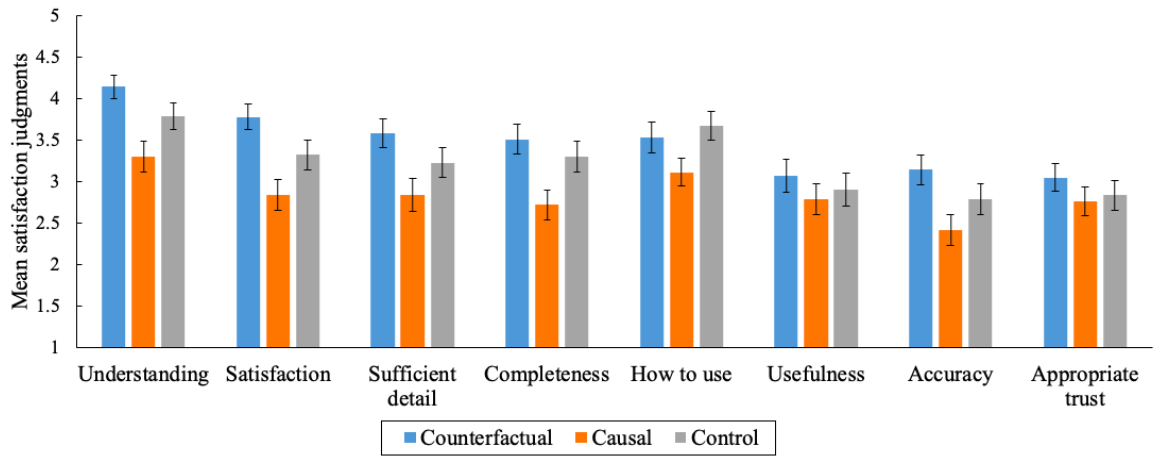

Figure 6: Mean judgments for Explanation Satisfaction Scale (error bars represent standard error of the mean; 1=Strongly Disagree; 5=Strongly Agree).

*Trust Measure.* A one-way ANOVA was carried out on the summed judgments for the Trust Scale to examine group

differences in trust levels for the explanations provided (see Figure 7). Significant differences between the groups were identified $F(2, 126)=8.184$, $p<.001$, $\eta_p^2=.117$. Post hoc Tukey HSD tests showed that the counterfactual group ($M=26.15$, $SD=6.14$) gave significantly higher trust judgments than the causal group ($M=20.21$, $SD=6.27$), $p<.001$, $d=.88$. The control group ($M=23.12$, $SD=7.63$) did not differ significantly from either the counterfactual ($p=.101$) or causal groups ($p=.115$). A reliable trend was identified when rank-ordering judgments for each item in the order Counterfactual > Control > Causal, $L(8)=112.0$, $p<.001$. Analogous to the satisfaction judgments, these results suggest that causal explanations engender less trust in comparison with counterfactual explanations and descriptions, while counterfactual explanations were deemed slightly more trustworthy than descriptions.

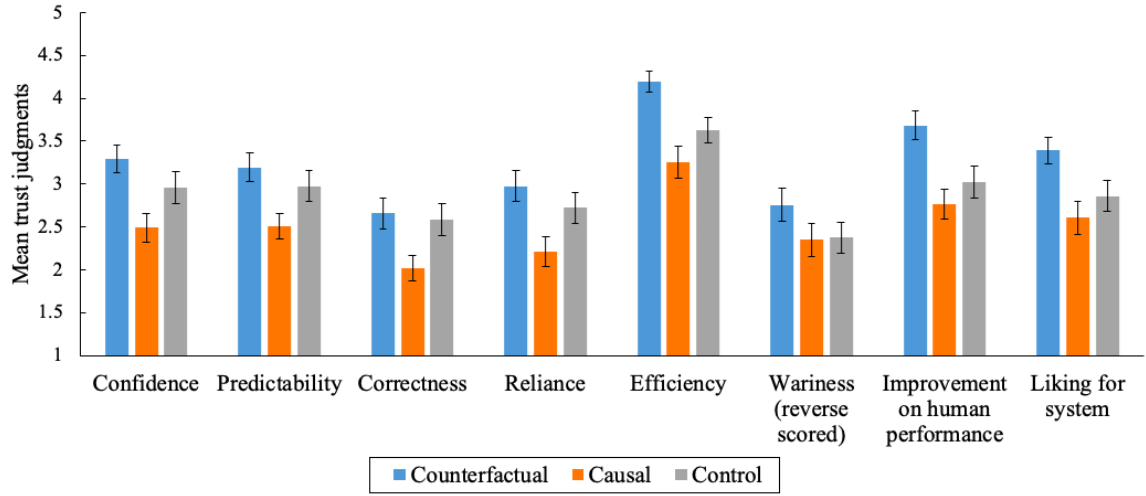

Figure 7: Mean judgments for Trust Scale (error bars represent standard error of the mean; 1=Strongly Disagree; 5=Strongly Agree).

## 5 GENERAL DISCUSSION

We have discovered that users' accuracy in predicting a system's decisions improves (i) when provided with explanations compared to none at all, with counterfactuals performing slightly better and being subjectively preferred compared to causal ones, and (ii) when using categorical features over continuous features, showing improvements over time between the training and test phases of the study. These novel results are significant for the light they shed on the application of explanations to elucidate the operations of an AI system. In the following sub-sections, we discuss their implications for (i) the role of explanations in XAI and the specific benefits that accrue to counterfactual explanations over causal explanations (and descriptions), (ii) the significance of categorical and continuous features in these explanations.

### 5.1 What Is It that (Counterfactual) Explanations Do?

There is an increasing recognition that explanations can play one of several roles in XAI. One of the main roles of an explanation is to improve the user's understanding of the domain, the AI system, or both, manifested by objective performance improvements in the task domain when explanations are provided. Measuring effects of explanation on objective performance is the guiding proposal in Hoffman's et al.'s [2018] conceptual framework for XAI and is a repeated theme in XAI user studies (see e.g., [6,36,63]). However, a number of studies indicate that explanations, especially counterfactual explanations, may *not* improve objective performance on the task (see [6,44,63]). On a related theme, many studies show that explanations are more likely to impact subjective assessments than objective performance; that is, people tend to self-report higher understanding [44] or judge decisions to be more fair [13], appropriate [3] or they express preferences for some explanation types in explanation-evaluation tasks [17]. These considerations combined with the present findings raise potentially serious ethical implications for use of explanations, as they suggest that some explanations may cause people to "feel better about" the AI system, without gaining any insight into why it made a prediction or how it works. Explanations may lead the explainee to a false assessment of the system, akin to the illusion of explanatory depth, wherein people overestimate their understanding of causal mechanisms underlying common phenomena [58], potentially leading to inappropriate trust of a system and its decisions.

The present results help to clarify the role that explanations may take. Overall, the counterfactual group were more accurate in both tasks than the control group, while the causal group's scores lay in between. This suggests that counterfactuals help people reason about the causal importance of the features used in the system's decisions more effectively than descriptions, and slightly better than causal explanations. Moreover, the counterfactual explanations improved people's accuracy in both phases without causing transfer or learning from the training and testing phase (i.e., there was an effect of Explanation, but this did not interact with any other factor). This conclusion is highly consistent with the key findings in the psychological literature where counterfactuals are shown to elicit causal reasoning and underlie people's understanding of causality [19,60]. Indeed, these findings also support proposals for the use of counterfactuals in algorithmic recourse [29,64], as they seem to better prompt an understanding of the predictions being made by the system. Notwithstanding, understanding a system's decision may only represent a first step towards recourse; as Barocas et al. [2020] highlight, counterfactual explanations should focus on features that reflect real world actions.

Furthermore, these objective benefits for the use of counterfactuals need to be set alongside the findings for their impacts on subjective measures. Here, we see that people are more satisfied and trusting in counterfactual explanations than causal explanations, and that the former are slightly preferred over descriptions. This result suggests that counterfactuals do not pose any greater threat to misleading subjective evaluations than simple descriptions of the predictions; one could conclude that the objective benefits of counterfactual explanations do not lead to inappropriate levels of satisfaction or trust. But, perhaps, the most surprising finding from the subjective measures is the low satisfaction and trust ratings for causal explanations relative to both counterfactuals and controls. This is inconsistent with findings in the XAI literature for causal rule-based explanation (e.g [36,63]) although it should be noted that the current experiment employs example-based causal explanations for each instance rather than general rule-based ones, which describe all the causal dependencies between features and an outcome. It may be the case that causal explanations prompt users to consider the relationship between feature and outcome more closely as evidenced by slightly higher accuracy compared to the control group, but these one-shot explanations do not provide any additional information to the user (unlike counterfactuals that suggest an alternative feature value), leading to lower subjective judgements. What is clear, is that, unlike counterfactual explanations, causal explanations do just marginally better than controls in accuracy, while attracting lower subjective evaluations.

### 5.2 The Primacy of Categorical Over Continuous Features

The results described here indicate that users were more accurate in making predictions based on categorical features than continuous features within each phase of the experiment. Furthermore, user accuracy increased in the testing phase relative to the training one, but this rise was mainly due to improvement in making predictions about categorical features (*gender* and *stomach-fullness*), one that does not occur for continuous features. As the three-way interaction is non-significant, we cannot attribute this effect to the provision of explanations; that is, it is an improvement that seems to emerge just from people gaining more experience throughout the training phase with the categorical features. Current counterfactual methods in XAI do not recognise any functional benefits for categorical features over continuous ones. These counterfactual methods transform categorical features to allow them to be processed similarly to continuous ones, using one-hot encoding or by mapping to ordinal feature spaces. So, no current model recognises that one feature-type might be more psychologically beneficial than another. Ironically, given the 100+ methods in the counterfactual XAI literature, no current algorithm gives primacy to categorical features over continuous ones in explaining the predictions of an AI system. Many models consider mutability and actionability as being important to the provided counterfactual explanations but neither of these concepts account for the results found here. Recall, the results show improved performance for the *gender* and *stomach-fullness* features even though the former is immutable and non-actionable (in the context of blood alcohol decisions) and the latter is mutable and actionable, as are *units* and *duration* (though *weight* in the context of blood alcohol decisions is immutable in the short term). Hence, the improvement in accuracy for these features over the course of the experiment (from training to testing phase) is more plausibly due to their simplicity (both have just two possible feature values) over the more complex continuous features (which have many possible feature values). There are clear implications here for counterfactual algorithms in algorithmic recourse; namely, that it would be better to focus on categorical features than on continuous ones when the predictive outcomes are equivalent.

### 5.3 Future Directions

The study and findings reported here give rise to several further lines of exploration. Firstly, the categorical features examined here are limited to binary values. Though these kinds of features commonly occur in many datasets (such as gender, ethnicity, or Boolean true/false features), categorical features can, in theory, have as many potential values as continuous ones. In this vein, an initial question may be whether there is a limit to the number of categorical values that humans can keep track of without compromising accuracy (or effectively, the categories become as challenging as continuous features). The hierarchy in accuracy observed between the features suggests that users may be able to monitor up to at least four categories, given than accuracy for *units* was higher than that for *weight* and *duration*, though further investigation is needed to validate this.

Finally, the study described here investigates differences between counterfactual and causal explanations, given the recent wave of interest in the former, suggesting that they are subjectively preferred to causal explanations, with some evidence that they aid in causal understanding of a system. Semi-factual explanations, which describe how some feature(s) could be perturbed towards the decision boundary without changing the class (e.g. "*Even if* you had asked for a lower loan of $10,000, your application would have been denied"), have also been proposed for XAI [35,53], although their explanatory potential has not been explored in controlled user tests as of yet. Given that semi-factuals can weaken causal relations drawn between antecedent and outcome [49], a comparison of their explanatory power to that of counterfactuals may be warranted.

### ACKNOWLEDGMENTS

This paper emanated from research funded by (i) the UCD Foundation and (ii) Science Foundation Ireland (SFI) to the Insight Centre for Data Analytics (12/RC/2289-P2).

# A APPENDICES

## A.1 Subjective Judgment Scales

### A.1.1 Explanation Satisfaction Scale Items

Table 1: Explanation Satisfaction Scale [24]

| Explanation Satisfaction Scale Items |
| --- |
| 1. From the explanation, I understand how the app works. |
| 2. The explanation of how the app works is satisfying. |
| 3. The explanation of how the app works has sufficient detail. |
| 4. This explanation of how the app works seems complete. |
| 5. This explanation of how the app tells me how to use it. |
| 6. This explanation of how the app works is useful to my goals. |
| 7. This explanation of the app shows me how accurate the app is. |
| 8. This explanation lets me judge when I should trust and not trust the app. |

### A.1.2 Trust Scale Items

Table 2: Trust Scale [24]

| Trust Scale Items |
| --- |
| 1. I am confident in the app. I feel that it works well. |
| 2. The outputs of the app are very predictable. |
| 3. The app is very reliable. I can count on it to be correct all the time. |
| 4. I feel safe that when I rely on the app I will get the right answers. |
| 5. The app is efficient in that it works very quickly. |
| 6. I am wary of the app. *(reverse scored)* |
| 7. The app can perform the task better than a novice human user. |
| 8. I like using the system for decision making. |

## A.2 Task Instructions

*Screen 1*

Thank you for agreeing to take part. Now, on to the experiment!

In this study, you are shown a new app, *SafeLimit*, that has been developed to inform people about when their blood-alcohol levels are over or under the legal limit to drive (i.e., 0.08% of total blood volume).

The app decides on a person's blood-alcohol content using information about:

| Gender | Male / Female |
|---|---|
| Weight | Weight of user (in kg) |
| Units | Amount of alcohol consumed (standard drinks) |
| Duration | Length of drinking period (in minutes) |
| Stomach | Full / Empty |

A unit of alcohol is equivalent to one standard drink (for example, a single shot measure of spirits, or a small glass of wine).

[Counterfactual condition: The app's decision is accompanied by an explanation telling you how changes in the person's details could change the decision about them being over or under the limit.]

[Causal condition: The app's decision is accompanied by an explanation telling you about the person's details that led to the decision about them being over or under the limit.]

You will see examples with details about different people and you are asked to judge whether that person is over or under the limit, before being shown the decision made by the *SafeLimit* app.

**What's Next?**

1. **Understanding the App's decisions:** In Part 1, you will be given some information about people's details in the *SafeLimit* app and you will be asked to judge whether each of these people is over or under the limit. [Counterfactual and Causal conditions: The *SafeLimit* app's decision and an explanation of its answer will then be shown.] [Control condition: The *SafeLimit* app's decision will then be shown]
2. **Predicting the App's decisions:** In Part 2, you will see more information about other people's details and your task is again to judge whether each person is over or under the limit. But you won't be shown the *SafeLimit* app's decision in this part.
3. **Your views on the App:** Finally, you will be asked some brief questions at the end about your views on the *SafeLimit* app overall.

*Screen 2*

Let's practice!

Here's an example of what you'll see in Part 1, Understanding the App's decisions. On each page, you will be shown the details of a person given to the *SafeLimit* app. In Part 1, you will see a total of 40 different examples of people's details. Your task, here, is to decide whether this person is over or under the legal limit.

If you think the person is over the limit, select the button that says "Over the limit".
If you think the person is under the limit, select the button that says "Under the limit".
If you don't know which of these options apply, select the button that says "Don't know".

Once you have made your selection, press the "Next" button.
Now, give it a try:

| Sarah | |
|---|---|
| Gender | Female |
| Weight | 64kg |
| Units | 4 |
| Duration | 90 mins |
| Stomach | Full |
| Limit | ? |

| Over the limit | Don't know | Under the limit |
|---|---|---|
|  |  |  |

Next

*Counterfactual condition*

*Screen 3: If correct response*

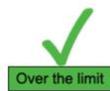

*Screen 3: If incorrect response*

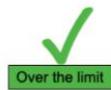
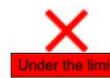

*Screen 3: If 'Don't know' response*

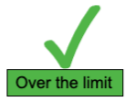
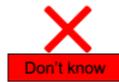
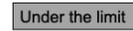

*Causal condition*

*Screen 3: If correct response*

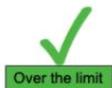
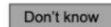
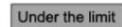

*Screen 3: If incorrect response*

You made an incorrect decision "Under the limit", which has turned red.
The correct decision is "Over the limit", which has turned green.

| Sarah | |
|---|---|
| Gender | Female |
| Weight | 64kg |
| **Units** | **4** |
| Duration | 90 mins |
| Stomach | Full |
| Limit | Over |

**Explanation**
Sarah is **over the limit** because she drank **4 units**.

Over the limit    Don't know     Under the limit

*Screen 3: If 'Don't know' response*

You made an incorrect decision "Don't know", which has turned red.
The correct decision is "Over the limit" which has turned green.

| Sarah | |
|---|---|
| Gender | Female |
| Weight | 64kg |
| **Units** | **4** |
| Duration | 90 mins |
| Stomach | Full |
| Limit | Over |

**Explanation**
Sarah is **over the limit** because she drank **4 units**.

Over the limit     Don't know    Under the limit

*Control condition*

*Screen 3: If correct response*

*Screen 3: If incorrect response*

*Screen 3: If 'Don't know' response*

You made an incorrect decision "Don't know", which has turned red.
The correct decision is "Over the limit" which has turned green.

| Sarah | |
|---|---|
| Gender | Female |
| Weight | 64kg |
| **Units** | **4** |
| Duration | 90 mins |
| Stomach | Full |
| Limit | Over |

Sarah is **over the limit**.

*Screen 4*

Let's Practice Part 2!

Now, we'll move on to the instructions for **Part 2, Predicting the App's decisions**.

When you finish the examples in Part 1, you will then be shown some more examples of information given to the *SafeLimit* app. Here in Part 2, you will see 40 more examples, which will be different to those you saw in Part 1. Again, your task here is to decide whether this person is over or under the legal limit, given a particular aspect of that person's details.

If you think the person is over the limit, select the button that says "Over the limit".
If you think the person is under the limit, select the button that says "Under the limit".
If you don't know which of these options apply, select the button that says "Don't know".

In Part 2, you will not be shown the app's decision.
Once you have made your selection, press the "Next" button.
**Now, let's practice again on this example**:
Given this person's gender, please make a judgment about their blood alcohol level.

| Max | |
|---|---|
| Gender | Male |
| Weight | 77kg |
| Units | 6 |
| Duration | 120 mins |
| Stomach | Empty |
| **Limit** | ? |

Over the limit    Don't know    Under the limit

Next

*Screen 5*

You have now finished reading the instructions for the study.

When you click the "Next" button below, the study will start and you will begin Part 1, **Understanding the App's decisions**.